\documentclass[reprint,prb,amssymb,amsmath,fleqn,floatfix]{revtex4-1}

\usepackage{graphicx,bm,dcolumn,textcomp}
\usepackage[hidelinks]{hyperref}

\begin{document}

\title[Relationship between mass density, electron density, and elemental composition]{Relationship between mass density, electron density, and elemental composition of body tissues for Monte Carlo simulation in radiation treatment planning}

\author{Nobuyuki~Kanematsu}\email{nkanemat@nirs.go.jp}
\affiliation{Department of Accelerator and Medical Physics, Research Center for Charged Particle Therapy, National Institute of Radiological Sciences, 4-9-1 Anagawa, Inage-ku, Chiba 263-8555, Japan}

\date{\today}

\begin{abstract}
\noindent{\bf Purpose:}
For Monte Carlo simulation of radiotherapy, x-ray CT number of every system needs to be calibrated and converted to mass density and elemental composition.
This study aims to formulate material properties of body tissues for practical two-step conversion from CT number.

\noindent{\bf Methods:}
We used the latest compilation on body tissues that constitute reference adult male and female.
We formulated the relations among mass, electron, and elemental densities into polylines to connect representative tissues, for which we took mass-weighted mean for the tissues in limited density regions.
We compared the polyline functions of mass density with a bi-line for electron density and broken lines for elemental densities, which were derived from preceding studies. 

\noindent{\bf Results:}
There was generally high correlation between mass density and the other densities except of C, N, and O for light spongiosa tissues occupying 1\% of body mass. 
The polylines fitted to the dominant tissues and were generally consistent with the bi-line and the broken lines.
 
\noindent{\bf Conclusions:}
We have formulated the invariant relations between mass and electron densities and from mass to elemental densities for body tissues. 
The formulation enables Monte Carlo simulation in treatment planning practice without additional burden with CT-number calibration.
\end{abstract}



\maketitle

\section{Introduction}

In treatment planning (TP) of radiotherapy, a patient body is commonly modeled to be H$_{2}$O of variable effective density, which corresponds to electron density for high-energy photons or stopping-power ratio for charged particles.
The effective density is normally converted from computed-tomography (CT) number for  attenuation of kilovoltage (kV) x ray.
The standard approach to construct such conversion functions involves experimental modeling of the x ray and stoichiometric analysis of body tissues.\cite{Schneider1996, Kanematsu2003}
The effective density is, though practically successful,\cite{Kanematsu2014,Inaniwa2015} an approximate concept for radiations undergoing complex interactions.

With the advancement of computer technology, Monte Carlo (MC) simulation of radiotherapy is becoming feasible, where a radiation is handled as a collection of particles individually interacting with matter of known composition according to the basic laws of physics.
For the modeling of body tissues, Schneider, Bortfeld, and Schlegel (SBS) applied the stoichiometric calibration to construct functions to convert CT number to mass density and elemental weights.\cite{Schneider2000} 
Their conversion functions, though commonly used for research,\cite{Paganetti2008,Mairani2013} are not applicable to the other CT systems.
Calibration and maintenance of the complex one-to-many relations may prevent MC simulation from applying to TP practice.

For patient dose calculation, TP systems commonly use a selectable function to convert CT number to effective density of interest.
Recently, a practical two-step approach was proposed for TP with proton and ion beams,\cite{Kanematsu2012,Farace2014} where CT number is only converted to electron density that is automatically converted to interaction-specific effective densities using invariant relations.
In the present study, we extend the two-step approach to promote MC-based TP practice. 

\section{Materials and Methods}

\subsection{Standard tissues and material properties}

We used the standard body tissue data in ICRP Publication 110,\cite{ICRP2009} which is the latest compilation of the kind.
In the publication are mass density $\rho$ and elemental weights $w$ for 53 standard tissues that fully comprise 141 organs of Reference Male and Female, for which we derived electron density $\rho_e$ and mass fraction $m$ per person (50\% male/50\% female).

Ignoring tiny mass of air, below 0.90 g/cm$^3$ was only a lung at 0.384 g/cm$^3$ occupying 1.4\% of body mass, which we chose for elemental composition of air-containing tissues.
In the 0.90--1.00 g/cm$^3$ region were an adipose tissue at 0.95 g/cm$^3$ (33.9\%) and medullary cavities including bone marrow all at 0.98 g/cm$^3$ (0.4\%).
In the 1.00--1.07 g/cm$^3$ region were a muscle at 1.05 g/cm$^3$ (34.5\%), spongiosa tissues (0.3\%), and many other general organs (11.7\%). 
In the 1.07--1.101 g/cm$^3$ region were a skin (4.9\%), a cartilage (0.6\%), and spongiosa tissues (0.7\%), which are miscellaneously epithelium, connective, and fatty bone tissues. 
In the 1.101--1.25 g/cm$^3$ region were only spongiosa tissues (5.7\%).
Above 1.25 g/cm$^3$ were a mineral bone at 1.92 g/cm$^3$ (5.7\%) and a tooth at 2.75 g/cm$^3$ (0.1\%). 

Among the material properties, we adopted mass density $\rho$ as the independent variable and examined its correlation to the dependent variables: electron density $\rho_e$ and elemental densities $\rho\,w$ of six major elements $M$ = \{H, C, N, O, P, Ca\}.
To define the regional representative tissues, we took mass-weighted mean for the set of densities $\vec{\rho} = (\rho, \rho_e, \rho\,w_\mathrm{H}, \rho\,w_\mathrm{C}, \rho\,w_\mathrm{N}, \rho\,w_\mathrm{O}, \rho\,w_\mathrm{P}, \rho\,w_\mathrm{Ca})$ over tissue $t$ in region $R$ by
\begin{eqnarray}
\vec{\rho}_R = \frac{\sum_{t \in R}{\vec{\rho}_t \, m_t}}{\sum_{t \in R}{m_t}}.
\end{eqnarray}
For the standard tissues and the regional representative tissues, we calculated the residual weight and the mean residual atomic number by adding weights and averaging atomic number $Z_r$ over residual element $r$, 
\begin{eqnarray}
w_\mathrm{res} = \sum_{r \notin M} w_r \quad \text{and} \quad
\bar{Z}_\mathrm{res} = \frac{\sum_{r \notin M}  Z_r w_r}{w_\mathrm{res}}, 
\end{eqnarray}
with which the residual mass could be approximately included in MC simulation.

\subsection{Density segmentation and tissue mixing}

Compared to muscle/organ tissues, adipose/marrow tissues have high concentration of fat.
Teeth have high concentration of minerals, connective tissues have high concentration of collagen, and bones are in between them. 
The concentration varies among and within individual tissues and are generally correlated with density.
Anatomically, adipose tissues are neighboring to muscles and organs, muscles are connected to bones via connective tissues, and teeth are connected to jaw bones.
At these interfaces, tissue mixing may occur in an image of finite spatial resolution.
Therefore, we modeled an arbitrary tissue of mass density $\rho$ to be mixture of representative tissues 1 and 2 that comprise a density segment of $\rho_1 \leq \rho \leq \rho_2$.
The other densities are interpolated from those of the representative tissues by
\begin{eqnarray}
\vec{\rho} = \frac{\rho_2-\rho}{\rho_2-\rho_1} \vec{\rho}_1+\frac{\rho-\rho_1}{\rho_2-\rho_1} \vec{\rho}_2 \label{eq:3}
\end{eqnarray}
in a mass-weighting manner.\cite{Schneider2000,Warren2015}
In other words, we assigned the representative tissues to the polyline points for conversion from mass density to the other densities.

For fatty tissues lighter than the representative adipose/marrow tissue, we extended the soft-tissue region down to 0.90 g/cm$^3$, which is the mass density of human fat at 37$^\circ$C.\cite{Fidanza2003}
Similarly, we extended the region for air-containing tissues up to 0.80 g/cm$^3$, leaving a transition segment of 0.80--0.90 g/cm$^3$ to avoid discontinuity of the material properties and to cope with body fat that escape over the segment boundary as viewed in CT image.
We also extended the hard-tissue region up to 3.00 g/cm$^3$ for teeth heavier than their representative.
To define the extended boundaries, we applied Eq.~\ref{eq:3} for extrapolation although, for the fat, the N, P, and Ca weights were forced to zero and the C weight was adjusted to sum up to 100\% to comply with general composition of fatty acid. 
For the polyline tissues, we compiled mass fraction, mass and electron densities, elemental and residual weights, and mean residual atomic number.

\subsection{Comparison with preceding formulations}

H\"unemohr {\it et al.} formulated a bi-line relation between mass and electron densities for the body tissues compiled in Refs.~\onlinecite{Woodard1986,White1987}.\cite{Huenemohr2014} 
From the same dataset, SBS selected their representative tissues for mixing of air/lung, adipose tissue/adrenal gland, small intestine/connective tissue, and marrow/cortical bone,\cite{Schneider2000} from which we derived their relations between mass density and elemental densities as broken-line functions.
We compared our results with those preceding formulations. 

\section{Results}

\begin{figure}
\includegraphics[width=8.5 cm]{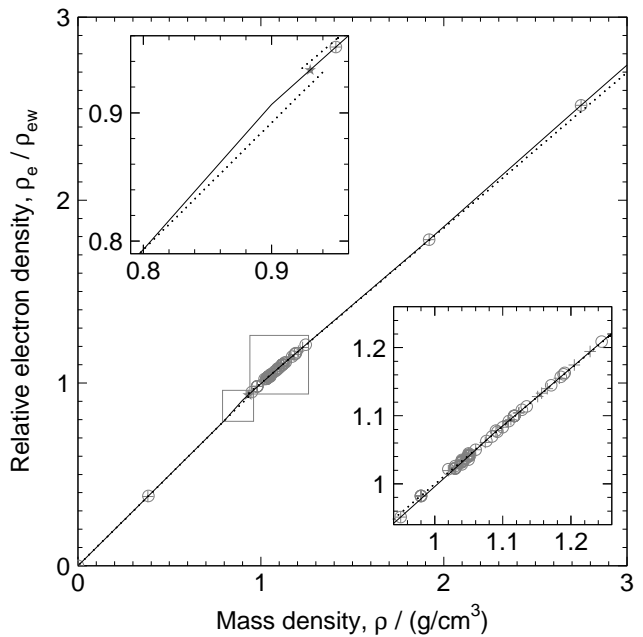}
\caption{Correlation between mass and electron densities for male ($+$) and female ($\bigcirc$) tissues plotted with the polyline function (solid lines), the bi-line function (dotted lines) and ``adipose 3'' ($\star$) at 0.93 g/cm$^3$ by H\"unemohr {\it et al.} (2014)\cite{Huenemohr2014}, and embedded subplots for the box-shaped areas.}
\label{fig:1}
\end{figure}

\begin{figure}
\includegraphics[width=8.5 cm]{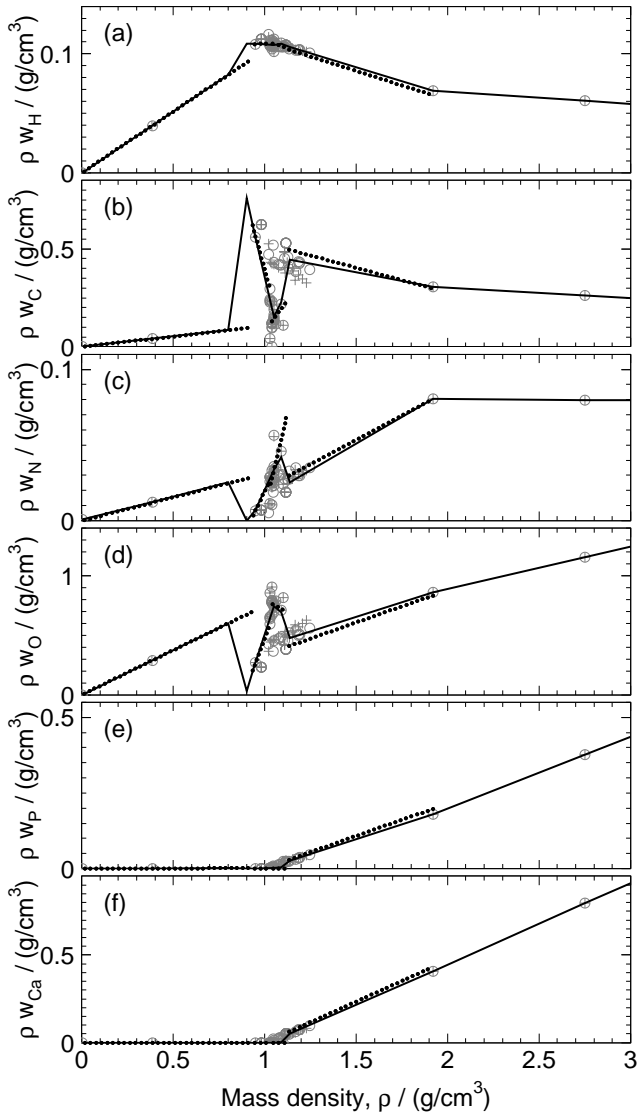}
\caption{Correlation between mass density and elemental densities of (a) H, (b) C, (c) N, (d) O, (e) P, and (f) Ca for male ($+$) and female ($\bigcirc$) tissues plotted with the polyline functions (solid lines) and the broken-line functions (dotted lines) according to Schneider, Bortfeld, and Schlegel (2000)\cite{Schneider2000}.}
\label{fig:2}
\end{figure}

\begin{figure}[b]
\includegraphics[width=8.5 cm]{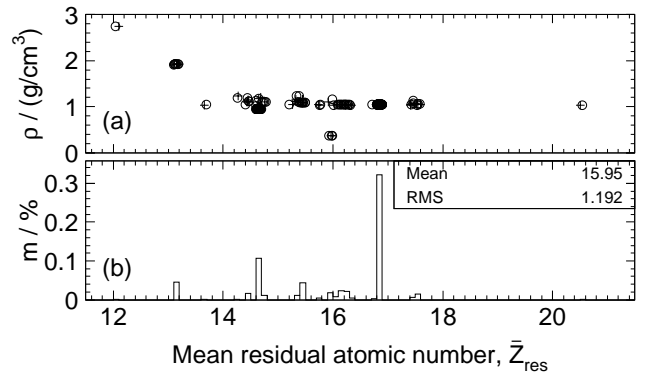}
\caption{(a) Scatter plot of mean residual atomic number versus mass density for male ($+$) and female ($\bigcirc$) tissues with thickness varied with residual mass and (b) mass-weighted histogram of mean residual atomic number.}
\label{fig:3}
\end{figure}

\begin{table*}
\caption{Material properties of the polyline tissues: mass fraction $m$, mass density $\rho$, relative electron density $\rho_e/{\rho_e}_\mathrm{w}$, elemental and residual weights $w_{\{\mathrm{H, C, N, O, P, Ca}\}}$ and $w_\mathrm{res}$, and mean residual atomic number $\bar{Z}_{\mathrm{res}}$, based on ISO Standard 2533 (1975)\cite{ISO1975} for the air and on ICRP Publication 110 (2009)\cite{ICRP2009} for the regional representative tissues.}
\begin{tabular}{rlddddddddddd}
\hline\hline 
\textnumero & 
Tissue type & 
\multicolumn{1}{c}{$\cfrac{m}{\%}$} &
\multicolumn{1}{r}{$\cfrac{\rho}{\mathrm{g}/\mathrm{cm}^3}$} & 
\multicolumn{1}{r}{$\cfrac{\rho_e}{{\rho_e}_\mathrm{w}}$} & 
\multicolumn{1}{r}{$\cfrac{w_\mathrm{H}}{\%}$} &
\multicolumn{1}{r}{$\cfrac{w_\mathrm{C}}{\%}$} &
\multicolumn{1}{r}{$\cfrac{w_\mathrm{N}}{\%}$} &
\multicolumn{1}{r}{$\cfrac{w_\mathrm{O}}{\%}$} &
\multicolumn{1}{r}{$\cfrac{w_\mathrm{P}}{\%}$} &
\multicolumn{1}{r}{$\cfrac{w_\mathrm{Ca}}{\%}$} &
\multicolumn{1}{r}{$\cfrac{w_\mathrm{res}}{\%}$} &
\multicolumn{1}{c}{$\bar{Z}_\mathrm{res}$} \\
\noalign{\smallskip}\hline\noalign{\smallskip}
1 & Air & & 0.001 & 0.001 & 0.00 & 0.01 & 75.52 & 23.17 & 0.00 & 0.00 & 1.30 & 18.0 \\
2 & Lung & 1.4 & 0.384 & 0.380 & 10.3 & 10.7 & 3.2 & 74.6 & 0.2 & 0.0 & 1.0 &15.9 \\  
3 & Air-containing & & 0.80 & 0.793 & 10.3 & 10.7 & 3.2 & 74.6 & 0.2 & 0.0 & 1.0 & 15.9 \\
\cline{3-3}
4 & Fat & & 0.90 & 0.907 & 12.09 & 84.39 & 0.00 & 3.52 & 0.00 & 0.00 & 0.00 &  \multicolumn{1}{c}{n/a} \\
5 & Adipose/marrow & 34.3 & 0.950 & 0.952 & 11.40 & 58.92 & 0.74 & 28.64 & 0.00 & 0.00 & 0.30 & 14.7 \\
6 & Muscle/organ & 46.5 & 1.049 & 1.040 & 10.25 & 14.58 & 3.20 & 70.87 & 0.21 & 0.02 & 0.87 & 16.8 \\
7 & Miscellaneous & 6.3 & 1.090 & 1.077 & 9.94 & 20.90 & 3.84 & 63.73 & 0.45 & 0.27 & 0.87 & 15.5 \\
8 & Spongiosa & 5.7 & 1.137 & 1.116 & 9.30 & 39.15 & 2.22 & 41.71 & 2.36 & 4.60 & 0.66 & 14.9 \\
9 & Mineral bone & 5.7 & 1.92 & 1.784 & 3.6 & 15.9 & 4.2 & 44.8 & 9.4 & 21.3 & 0.8 & 13.1 \\
10 & Tooth & 0.1 & 2.75 & 2.518 & 2.2 & 9.5 & 2.9 & 42.1 & 13.7 & 28.9 & 0.7 & 12.0 \\
11 & Extra tooth & & 3.00 & 2.739 & 1.93 & 8.27 & 2.65 & 41.58 & 14.53 & 30.37 & 0.67 & 11.8 \\
\hline\hline 
\end{tabular}
\label{tab:1}
\end{table*}

Figure~\ref{fig:1} shows the correlation between mass and electron densities, where ${\rho_e}_\mathrm{w} = 3.343 \times 10^{23}/\mathrm{cm}^3$ is the electron density of water.
The two fitting functions, the polyline and the bi-line by H\"unemohr {\it et al.}, were generally consistent with the standard tissues except around the fat at 0.90 g/cm$^3$ and around the tooth at 2.75 g/cm$^3$, for which the bi-line function gave 0.892 ($- 1.7\%$) and 2.486 ($-1.3\%$), respectively.
The discontinuity in the bi-line at their ``adipose 3'' was settled by the polyline with better fitting.
Figure~\ref{fig:2} shows the correlation between mass density and elemental densities.
The polyline functions and the broken-line functions according to SBS were both generally consistent with the standard tissues except for the C, N, and O densities in the 1.0--1.1 g/cm$^3$ region.
As shown in Fig.~\ref{fig:3}, the highest and lowest mean residual atomic numbers were 20.6 at 1.04 g/cm$^3$ for the thyroid and 12.0 at 2.75 g/cm$^3$ for the tooth.
The global mean of 15.95 approximately corresponded to element S.
Table~\ref{tab:1} shows the resultant material properties of the polyline tissues.

\section{Discussion}

A concise yet complete dataset of the polyline tissues may be suited to stoichiometric analysis for CT-number calibration.\cite{Schneider1996,Kanematsu2003}
In the cases where electron density is available, the most likely mass and elemental densities can be determined by Eq.~\ref{eq:3} with variables $(\rho, \rho_1, \rho_2)$ replaced to $(\rho_e, {\rho_e}_1, {\rho_e}_2)$.
The resultant mass density and composition including residual weight for element S will constitute a volumetric patient model for MC simulation. 

The general agreement between the formulations indicates the consistency between the compilations of body-tissue data. 
The mass-weighting approach of this study took advantage of the publication that focused on the computational phantoms of reference adult male and female rather than on variations in age, physical status, or individual.\cite{ICRP2009}

The poor fitting in C, N, and O densities was due to undifferentiated inclusion of spongiosa tissues occupying 1\% of body mass in the 10.--1.1 g/cm$^3$ region.
They could potentially be resolved by anatomical identification or independent quantitative imaging, in which case either an extended spongiosa/mineral-bone segment or a separate marrow/spongiosa segment should be applied to them. 

Recently, megavoltage (MV) CT and dual-energy (DE) CT have been investigated for direct electron-density imaging.\cite{Ruchala2000,Landry2011}
H\"unemohr {\it et al.} formulated the composition of body tissues as a two-variable function of electron density and effective atomic number from DECT with improved accuracy.\cite{Huenemohr2014}
The DECT with kV and  MV x rays would be ideal.\cite{Bazalova2008}
Nevertheless, use of MVCT or DECT will potentially mitigate metal artifact and beam hardening, which limits the accuracy of kVCT.\cite{Wu2014,Yu2012} 


In conclusion, we have formulated the invariant polyline relations between mass and electron densities and from mass to elemental densities for body tissues. 
The formulation enables MC simulation in TP practice without additional burden with CT-number calibration.


The author thanks M.~Nakao of HIPRAC, Hiroshima for private communication on his inspiring research.


\end{document}